\documentclass[aps,prl,twocolumn, superscriptaddress, showpacs]{revtex4-1}

\usepackage{amsmath,amssymb}
\usepackage{xcolor}
\usepackage{graphicx}
\usepackage{braket}
\usepackage[utf8]{inputenc}

\begin{document}

\newcommand{\h}[1]{\hat{#1}}
\newcommand{\sinc}{\mathrm{sinc}}
\newcommand{\todo}[1]{\textcolor{blue}{ToDo: {#1}}}
\newcommand{\BS}{boson-sampling }

\newcommand{\tim}{Applied Physics, University of Paderborn, Warburger Stra\ss e 100, 33098 Paderborn, Germany}
\newcommand{\prague}{FNSPE, Czech Technical University in Prague, Br\^ehov\'{a} 7, 119 15, Praha 1, Czech Republic}

\title{Driven Boson Sampling}

\author{Sonja Barkhofen}
\email{Corresponding author: sonja.barkhofen@upb.de}
\affiliation{ \tim}
\author{Tim J. Bartley}
\author{Linda Sansoni}
\author{Regina Kruse}
\author{Craig S. Hamilton}
\affiliation{\prague}
\author{Igor Jex}
\affiliation{\prague}
\author{Christine Silberhorn}
\affiliation{ \tim}

\begin{abstract}
Sampling the distribution of bosons that have undergone a random unitary evolution is strongly believed to be a computationally hard problem. 
Key to outperforming classical simulations of this task is to increase both the number of input photons and the size of the network.
We propose driven boson sampling, in which photons are input within the network itself, as a means to approach this goal.  
When using heralded single-photon sources based on parametric down-conversion, this approach offers an $\sim e$-fold enhancement in the input state generation rate over scattershot boson sampling, reaching the scaling limit for such sources.
More significantly, this approach offers a dramatic increase in the signal-to-noise ratio with respect to higher-order photon generation from such probabilistic sources, which removes the need for photon number resolution during the heralding process as the size of the system increases.
\end{abstract}

\maketitle

\begin{figure}
 \centering
 \includegraphics[width=0.8\columnwidth]{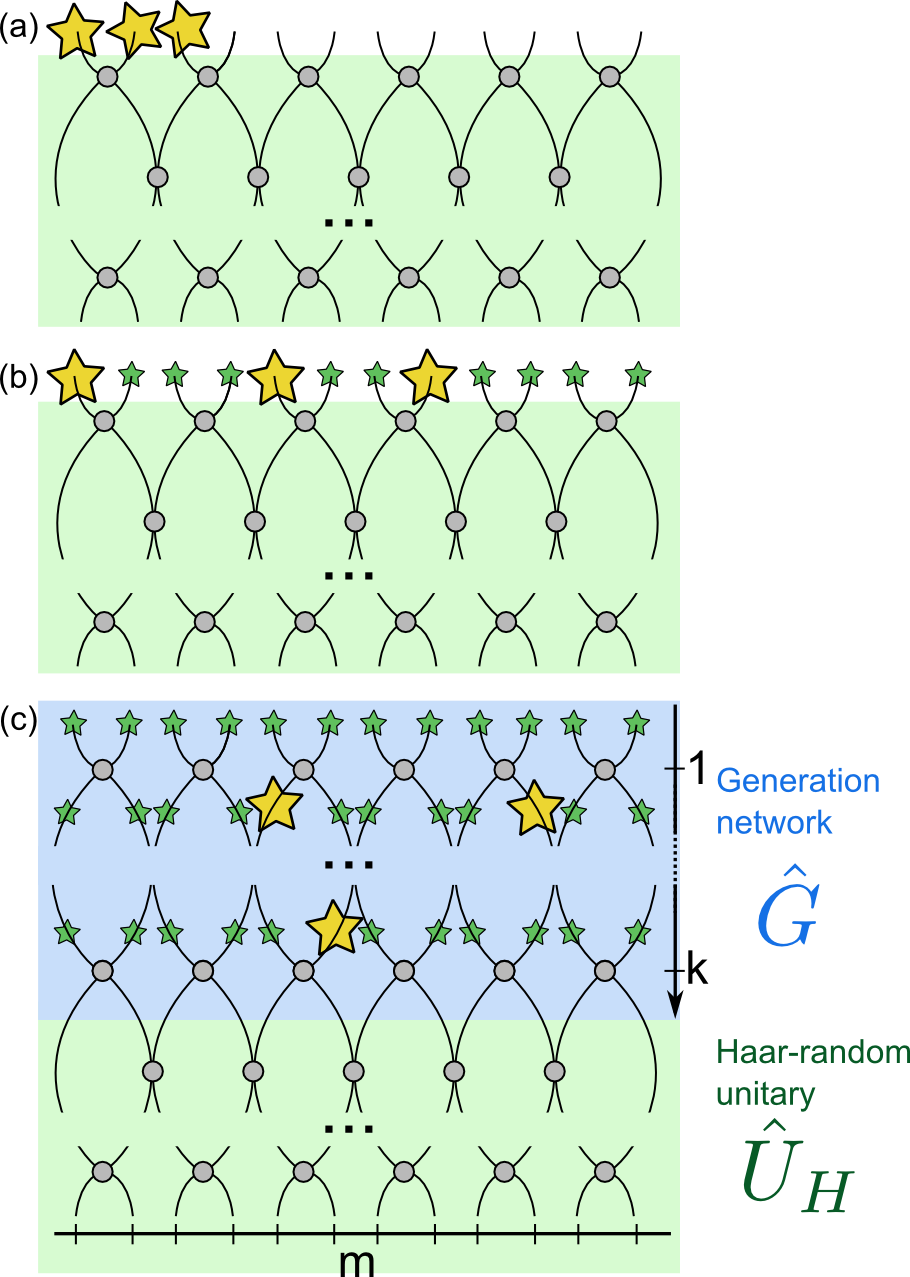}
 \caption{\label{fig:BS_network} Boson sampling networks (a) for boson sampling; (b) for scattershot boson sampling; (c) for driven boson sampling with the state generation part governed by the $k\cdot m\times m$ matrix $\hat{\mathbf{G}}$ followed by a Haar-random network $\hat{\mathbf{U}}_H$; green stars indicate all possible modes for injecting photons, yellow stars mark one possible $(n=3)$-input state.}
\end{figure}

%
Boson sampling~\cite{aaronson_computational_2011} is strongly conjectured to be a computationally hard problem.
It describes the sampling from the output distribution of indistinguishable bosons evolving through a sufficiently large random unitary, as depicted in Fig.~\ref{fig:BS_network} (a).
Though it is not a universal quantum computation problem, boson sampling has attracted considerable attention due to its experimental feasibility with quantum optics. 
Different photonic platforms have demonstrated inputting up to 4 single photons in networks of up to 13 input modes~\cite{broome_photonic_2013,spring_boson_2013,tillmann_experimental_2013,crespi_integrated_2013,spagnolo_experimental_2014,bentivegna_experimental_2015, carolan_universal_2015,loredo_bosonsampling_2016,he_scalable_2016}.
However, it remains a challenge to scale up the devices to $20$-$30$ photons \cite{aaronson_computational_2011} traversing a correspondingly large network, a regime in which a quantum boson sampling machine is expected to outperform classical computers.

In the first boson sampling experiments~\cite{broome_photonic_2013,spring_boson_2013,tillmann_experimental_2013,crespi_integrated_2013}, parametric down conversion (PDC) sources were employed and thus the photons were generated in a probabilistic fashion.
With this scheme, measurement time scales exponentially with photon number.
To improve this performance, two complementary approaches have been developed. 
Recently, source hardware has been improved by implementing quasi-on-demand single photon sources as inputs to a boson sampling circuit~\cite{loredo_bosonsampling_2016,he_scalable_2016}, resulting in a significant reduction in measurement time. 
In parallel, algorithmic (``software'') developments have also improved the scaling when using probabilistic sources. Scattershot boson sampling (SBS)~\cite{lund_boson_2014, scott_aaronson_http://www.scottaaronson.com/blog/?p1579_????,bentivegna_experimental_2015} increases the number of possible inputs to the linear network, as shown in Fig.~\ref{fig:BS_network} (b), by a binomial factor, which reduces the measurement time by a corresponding amount. However, probabilistic PDC sources typically suffer from the additional limitation of high-order photon contributions. In general, as the required number of photons increases, the chance of higher-order terms also increases. 

If the number of {\em possible} inputs can be increased compared
to the required number of photons, the rate is increased while the effect of higher order terms arising from PDC can be reduced. 
This is because the pump power of each source can be reduced without lowering the overall generation probability.
In SBS, the number of input modes defines one dimension of the network, which in turn determines the required depth of the network. 
Therefore arbitrarily increasing the number of possible inputs necessarily increases the network size in both dimensions, which squares the number of required components. Thus the question arises: can the number of possible inputs be decoupled from the network dimension, such that sufficiently many possible inputs can be constructed without blowing up the network size?

To answer this question, we propose Driven Boson Sampling (DBS) as an experimentally accessible means to increase the number of possible inputs when using heralded PDC photon sources, independent of the size of the network. This approach both increases the input state generation rate and significantly reduces the effect of higher-order photon contributions, overcoming the need for photon-number resolving herald detectors.
In our scheme, we consider two connected networks of beam splitters, as shown in Fig.~\ref{fig:BS_network} (c). The underlying structure of each realises a unitary drawn from a Haar-random distribution. However, in the first network photons can be injected not just at the first layer but at all links between nodes during the evolution. This generation network consists of $k$ layers each comprising $\tfrac{m}{2}$ beam splitters. 
Evolution of $n$ photons through the generation network is therefore governed by the $k\cdot m\times m$ scattering matrix $\mathbf{G}$. 

This generation network serves as an input of the second $m\times m$ Haar-random unitary $\mathbf{U}_H$ network of width $m$ inputs and sufficient depth. 
In general, the requirements on size imply $m\gg n^2$ in order to reduce the chance of multiple photons in the output modes (overcoming the so-called birthday problem~\cite{aaronson_computational_2011}). 
It has been shown that, in the case of exact Boson sampling, the depth of this second network $\mathbf{U}_H$ need not exceed 4 layers of beam splitters to be classically hard~\cite{brod_complexity_2015}. However, for the case of approximate Boson sampling, the minimum depth bounds are $\mathcal{O}\left(n\log m\right)$~\cite{aaronson_computational_2011} and $\mathcal{O}\left(m\log m\right)$~\cite{scott_aaronson_http://www.scottaaronson.com/blog/?p1579_????} for standard and scattershot boson sampling, respectively. Since the simplest case of DBS has all photons generated in the final layer of the generation network $G$, this corresponds to SBS if $U_H$ is of depth $\mathcal{O}\left(m\log m\right)$. This provides a useful upper bound on the minimum depth of $U_H$. In principle, the minimum number of layers $k$ of $G$ is 1, since this corresponds exactly to SBS. However, when considering noise caused by higher-order photon contributions, we find that the minimum number of layers when $m=n^2$ is $k\geq\tfrac{1}{n\left(\sqrt[n]{2}-1\right)}$ in order to achieve a SNR$>1$ (see supplemental material for more details). We leave as an open question the computational complexity of sampling the dynamics governed by $\mathbf{G}$ alone, and the associated minimum depth requirements. 

\begin{figure}
 \centering
 \includegraphics[width=\columnwidth]{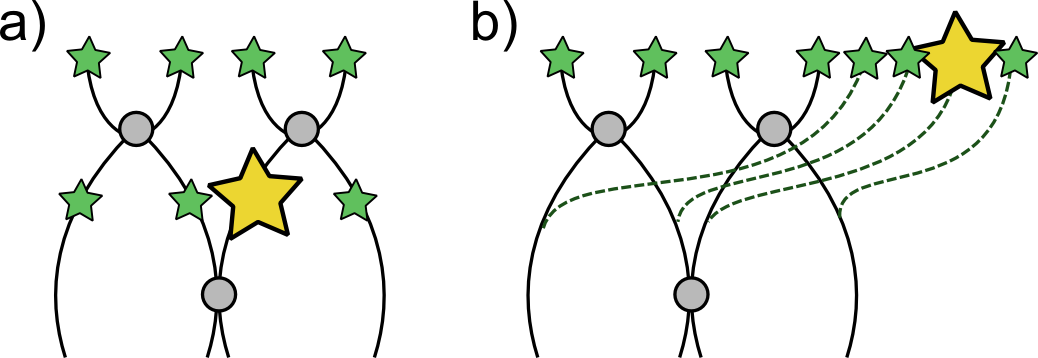}
 \caption{\label{fig:DBS_state} a) State generation in driven boson sampling; b) equivalent system with an adapted graph (by green dashed lines) and single photon input state only from the top; green stars indicate all possible modes for injecting photons, the yellow star marks an equivalent input position.}
\end{figure}

At first sight, it appears that by injecting bosons within a network we move away from the fundamental constraint of unitary dynamics to the nonlinear regime, which is not covered under existing hardness conjectures of boson sampling. However, this scheme can in fact be mapped to a valid boson sampling problem.
To illustrate, we begin with the abstraction shown in Fig.~\ref{fig:DBS_state}.
The input modes of each fundamental unit can be extended to the top of the network [as shown in Fig.~\ref{fig:DBS_state} (b)], creating an input state vector of length $k\cdot m$ at the top of the network, similar to the SBS case. We can then write the evolution of the whole system as a transformation of an input state of length $k\cdot m$ to an output state of length $m$, via the $k\cdot m\times m$ scattering matrix $\mathbf{G}$ and unitary evolution through the network governed by $\mathbf{U}_H$. 

\begin{figure}
 \centering
 \includegraphics[width=1\columnwidth]{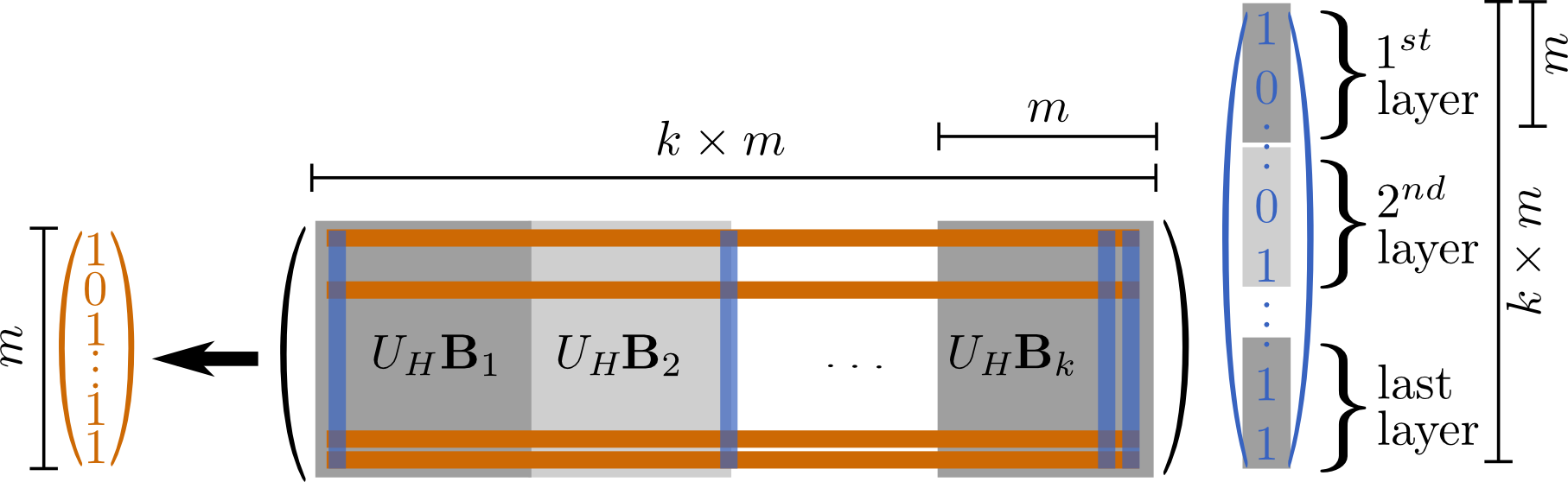}
 \caption{\label{fig:equationG} Pictorial relationship of the input state to a particular measurement outcome $\ket{S_\textrm{out}}$ due to matrix $\mathbf{G}$ built from $k$ blocks $B_i$ of size $m\times m$~\cite{spring_single_2014}.}
\end{figure}

The input state is 
\begin{equation}
\ket{S_\textrm{in}} = \bigotimes_{i = 1}^{k\cdot m}(a_i^\dag)^{s_i}\ket{0}_i=\ket{s_1,\ldots,s_{k\cdot m}} ~,
\label{eq:inputstate}
\end{equation}
where $a_i^\dag$ is a bosonic creation operator in mode $i$ and $s_i\in \left\{0,1\right\}$ describes single photons in $n$ of the $k\cdot m$ modes and vacuum otherwise.
After the evolution governed by $\mathbf{U}_H\mathbf{G}$ and a projective measurement, the measurement outcome $\ket{S_\textrm{out}}=\ket{t_1,\ldots,t_m}$ with $t_i\in \left\{0,1\right\}$ is related to the permanent of an $n\times n$ submatrix $\left[\mathbf{U}_H\mathbf{G}\right]^{(S_\mathrm{out}|S_\mathrm{in})}$ (the elements of which may be visualized by the intersection of the orange rows and the blue columns in Fig~\ref{fig:equationG}), following the procedure in {e.g.} Ref.~\cite{spring_single_2014}, such that the probability of a particular outcome $\ket{S_\mathrm{out}}$ given an input state $\ket{S_\mathrm{in}}$ is related to the permanent by 
\begin{equation}
P(S_\mathrm{out}|S_\mathrm{in}) \propto |\mathrm{Per}(\left[\mathbf{U}_H\mathbf{G}\right]^{(S_\mathrm{out}|S_\mathrm{in})})|^2~.
\label{eq:permanent}
\end{equation}
While it is long understood that calculating permanents of matrices is hard~\cite{valiant_complexity_1979}, the insight from Aaronson and Arkhipov in their original statement was to show that efficient \textit{sampling} from distributions governed by the permanents of $n\times n$ Gaussian matrices contained within an $n\times m$ scattering matrix would have profound implications for the hierarchy of computational complexity.
It is therefore strongly conjectured to be a $\#\textsf{P}$-hard problem (the permanent-of-Gaussians conjecture~\cite{aaronson_computational_2011}), even in the approximate case where we allow for errors. 
Building on this result, SBS~\cite{lund_boson_2014} extends the size of the scattering matrix to $m\times m$, and samples an ensemble average of $n$ photons in all possible $m$ inputs, which yields an $\left(\begin{array}{c}m\\n\end{array}\right)$ increase in the input state generation rate. 
In DBS, the scattering matrix is now of size $k\cdot m\times m$, yielding an enhancement input state generation proportional to $\left(\begin{array}{c}k\cdot m\\n\end{array}\right)$. 

To provide strong evidence for the complexity of this problem we show that the $n\times n$ submatrices which govern the evolution of a single instance of this DBS machine remain close in variation distance to a matrix of independent and identically distributed (i.i.d.) Gaussians, in line with theorem 3 of the original hardness conjecture~\cite{aaronson_computational_2011}. 
To do so, we must consider the form of each of the submatrices $\left[\mathbf{U}_H\mathbf{G}\right]^{(S_\mathrm{out}|S_\mathrm{in})}$. To build $\mathbf{G}$, we consider first the $m\times m$ unitary coupling matrices $\mathbf{C}_i$, which govern the pairwise interaction between inputs at layer $i$ (see Fig.~\ref{fig:BS_network}). The output modes of this interaction then become the inputs in layer $i+1$, and the evolution continues. These coupling matrices take the following form:
\begin{align}
\mathbf{C}_i=&
\left(\begin{array}{ccccccc}
1&&&&&&\\
&t_{i1}&r_{i1}&&&&\\
&-r_{i1}&t_{i1}&&&&\\
&&&\ddots&&&\\
&&&&t_{i\tilde{m}}&r_{i\tilde{m}}&\\
&&&&-r_{i\tilde{m}}&t_{i\tilde{m}}&\\
&&&&&&1\end{array}\right)~\textrm{for even $i$}~,\\
\mathbf{C}_k=&\left(\begin{array}{ccccc}
t_{l1}&r_{l1}&&&\\
-r_{l1}&t_{l1}&&&\\
&&\ddots&&\\
&&&t_{l\tilde{m}^\prime}&r_{l\tilde{m}^\prime}\\
&&&-r_{l\tilde{m}^\prime}&t_{l\tilde{m}^\prime}
\end{array}\right)~\textrm{for odd $l$}~,
\end{align}
where $\tilde{m}=\tfrac{(m-2)}{2}$, $\tilde{m}^\prime=\tfrac{m}{2}$, and the elements $t,r$ are the (real) beam splitter parameters satisfying $t^2+r^2=1$.

Each of the $m\times m$ blocks of the operator $\mathbf{G}$ (see grey blocks in Fig.~\ref{fig:equationG}) can be built from products of the coupling matrices. The $q$th block, $\mathbf{B}_q$ for $1\leq q \leq k$ which governs the evolution of the photons generated between layer $q-1$ and $q$, is given by the product of coupling matrices for subsequent layers $q,q+1,\ldots,k$, such that $\mathbf{B}_q = \prod_{i = q}^{k} \mathbf{C}_i$.
Note that the evolution through the first $1,2,\ldots, q-1$ layers is given by the identity (as depicted in Fig.~\ref{fig:DBS_state} (b)).

Although the matrix $\mathbf{G}$ is not a unitary operator, each constituent block $\mathbf{B}_q$ is unitary. Furthermore, since the evolution is governed by $\mathbf{U}_H\mathbf{G}$, we can write each block of the entire evolution matrix as $\mathbf{U}_H\mathbf{B}_q$. These blocks are again unitary and, as the Haar measure is invariant under the action of the underlying $U(m)$ group, each block is itself a random unitary according to the Haar measure. 

Thus, our $k\cdot m\times m$ evolution matrix $\mathbf{U}_H\mathbf{G}$ comprises $k$ blocks of $m\times m$ Haar-random unitaries, as shown in Fig.~\ref{fig:equationG}. Each instance of evolution through this network is governed by an $n\times n$ sub-matrix $\left[\mathbf{U}_H\mathbf{G}\right]^{(S_\mathrm{out}|S_\mathrm{in})}$ contained within $\mathbf{U}_H\mathbf{G}$. 
The essence of Aaronson and Arkhipov's complexity conjecture in Ref.~\cite{aaronson_computational_2011} is that a sufficiently small sample of elements from a Haar-random matrix contains insufficient structure to efficiently compute the permanent, {i.e.} that those elements are close in variation distance to a matrix of i.i.d. Gaussian elements. The submatrix sampled by our photons can comprise elements from different Haar-random blocks, therefore the elements of this matrix are at least as independent as elements sampled from a single $m\times m$ Haar-random unitary. Thus sampling the probability distribution arising from these submatrices retains at least the level of complexity as the original problem. 
Moreover, the complexity proofs for sampling from an ensemble of these matrices ({i.e.} SBS), must also apply in this case. 

We note also that the reduction in the opposite direction is straightforward: each instance of a DBS experiment in which photons are generated in the same layer of the generation network corresponds to a SBS problem. If one were able to efficiently compute the outcome of a DBS problem, then one could therefore efficiently extract a subset of the results corresponding to SBS.

\begin{figure}
 \centering
 \includegraphics[width=0.9\columnwidth]{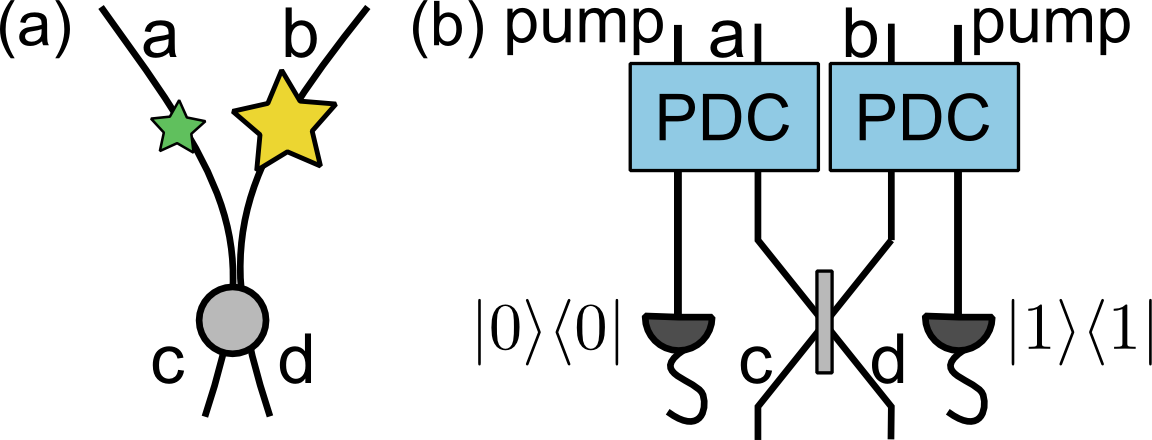}
 \caption{\label{fig:DBS_incoupling} Fundamental unit of the  network (a) photon creation at a link $b$ and no photon created at link $a$; (b) experimental implementation using parametric down conversion to generate a heralded photon.}
\end{figure}
To demonstrate the benefits of our scheme, we consider an experimental approach which is readily implemented using heralded parametric down-conversion (PDC), as shown in Fig.~\ref{fig:DBS_incoupling} (b). Measuring a single photon heralds the presence of a new photon within the network~\footnote{A linear optics implementation using measurement-induced nonlinearity is possible in principle, however very challenging experimentally. See Supplementary Material for further details.}. This process 
has been used extensively in photon addition experiments (see {e.g.}~\cite{zavatta_quantum--classical_2004,parigi_probing_2007} and supplementary material). 
Adding photons in this manner increases the total number of input modes in $\mathbf{G}$ to $k\cdot m$. It is necessary that each source can be heralded, such that it is known within each trial how many sources fire. Note that the timing of the additional inputs must be determined such that all photons have the potential to interact, regardless of the layer of the network in which they are generated.

\begin{figure}
 \centering
 \includegraphics[width=0.8\columnwidth]{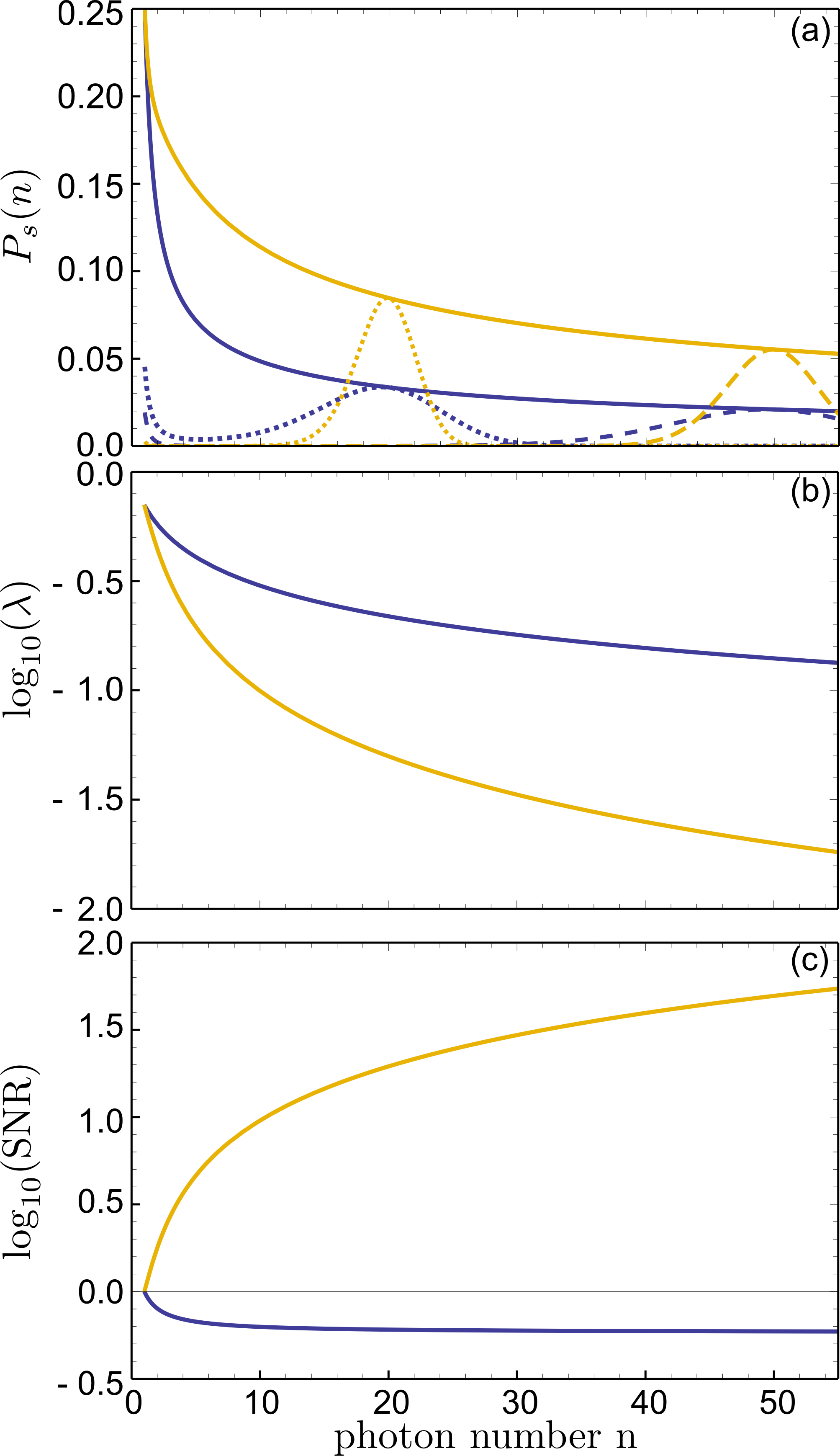}
 \caption{\label{fig:probs} Comparison of DBS (orange lines) and SBS (blue lines) as a function of photon number: (a) success probabilities $P_s(n)$ for optimal generation efficiencies $\lambda$ (solid lines) as well as two example distributions of $P_s(n)$ for optimized for $n = 20$ (dotted lines) and $n=50$ (dashed lines); (b) optimal generation efficiency $\lambda$ for an $n$ photon event; (c) signal-to-noise-ratio (SNR) of heralding and generating single photon events divided by the probability of higher order contributions (see supplementary material for details). The DBS case corresponds to $k=\sqrt{m}$, and the SBS corresponds to $k=1$. In both cases, we assume $m=n^2$.}
\end{figure}

In the original boson sampling scheme, single photons are input into the network in predetermined input modes, specifying a single configuration of modes with and without photons. If one is using $n$ heralded single photon sources at the input, for example arising from PDC, one must wait for all $n$ heralds before a boson sampling experiment can take place. This occurs with probability $P^{\mathrm{BS}}_s(n) = P_1^n$, where $P_1$ is the single-photon generation probability for a single source. In SBS, all $m$ input modes are coupled to heralded single photon sources. However, all possible configurations of exactly $n$ of the $m$ sources firing is a valid input state, therefore one gains an $m$ choose $n$ speed-up in the number of valid trials of a boson sampling experiment, whereby $P^{\mathrm{SBS}}_s(n) =\left(\begin{array}{c}m\\n\end{array}\right)P_1^nP_0^{m-n}$.
Here, $P_0$ is the probability of no photons (vacuum) being generated. 
In DBS, a valid generation event of $n$ single photons occurs with success probability $P^{\mathrm{DBS}}_s(n) =\left(\begin{array}{c}k\cdot m\\n\end{array}\right)P_1^nP_0^{k\cdot m-n}$, where $k\cdot m$ is the number of possible input positions. 

The advantage offered by DBS can be demonstrated by optimizing the single photon generation probability for a desired photon number $n$. 
For PDC states of the form $\ket{\psi_{\mathrm{PDC}} }= \sqrt{1-\lambda^2}\displaystyle\sum_{i=0}^\infty\lambda^i\ket{i,i}$ with $i$ the photon number, the probability of generating a photon is $P_1 = (1-\lambda^2)\lambda^2$, and vacuum $P_0 = (1-\lambda^2)$, where $\lambda$ is the squeezing parameter.  For fixed photon number $n$, and number of possible inputs $k\cdot m$, we can find the optimal $\lambda$ to maximize success probability $P_s(n)$:
\begin{equation}
\lambda_\textrm{opt}=\sqrt{\frac{n}{k\cdot m+n}}~.
\end{equation}
In principle, in DBS we are free to choose the number of layers $k$ under the constraint $k\cdot m\geq n^2$. Note that $k=1$ is the case for SBS. 
Following the supplemental material of Ref.~\cite{lund_boson_2014}, in the asymptotic limit for $m\geq n^2$, the scaling of optimal generation probability is $P_\textrm{max}\sim\frac{1}{b\sqrt{2\pi}}\frac{1}{\sqrt{n}}$, where the factor $b=e$ if $m=n^2$ and  $b=1$ otherwise. 
By choosing $k=\sqrt{m}=n$ which corresponds to $n^3$ input modes, the geometry of DBS allows the success probability $P_s(n)$ to be a factor of $e$ higher compared to SBS [Fig.~\ref{fig:probs} (a)], enabling more than twice the data-rate of scattershot boson sampling for fixed laser repetition rate (or equivalently, more than twice the acquired data for a fixed experiment time). 

\begin{figure}
 \centering
 \includegraphics[width=0.6\columnwidth]{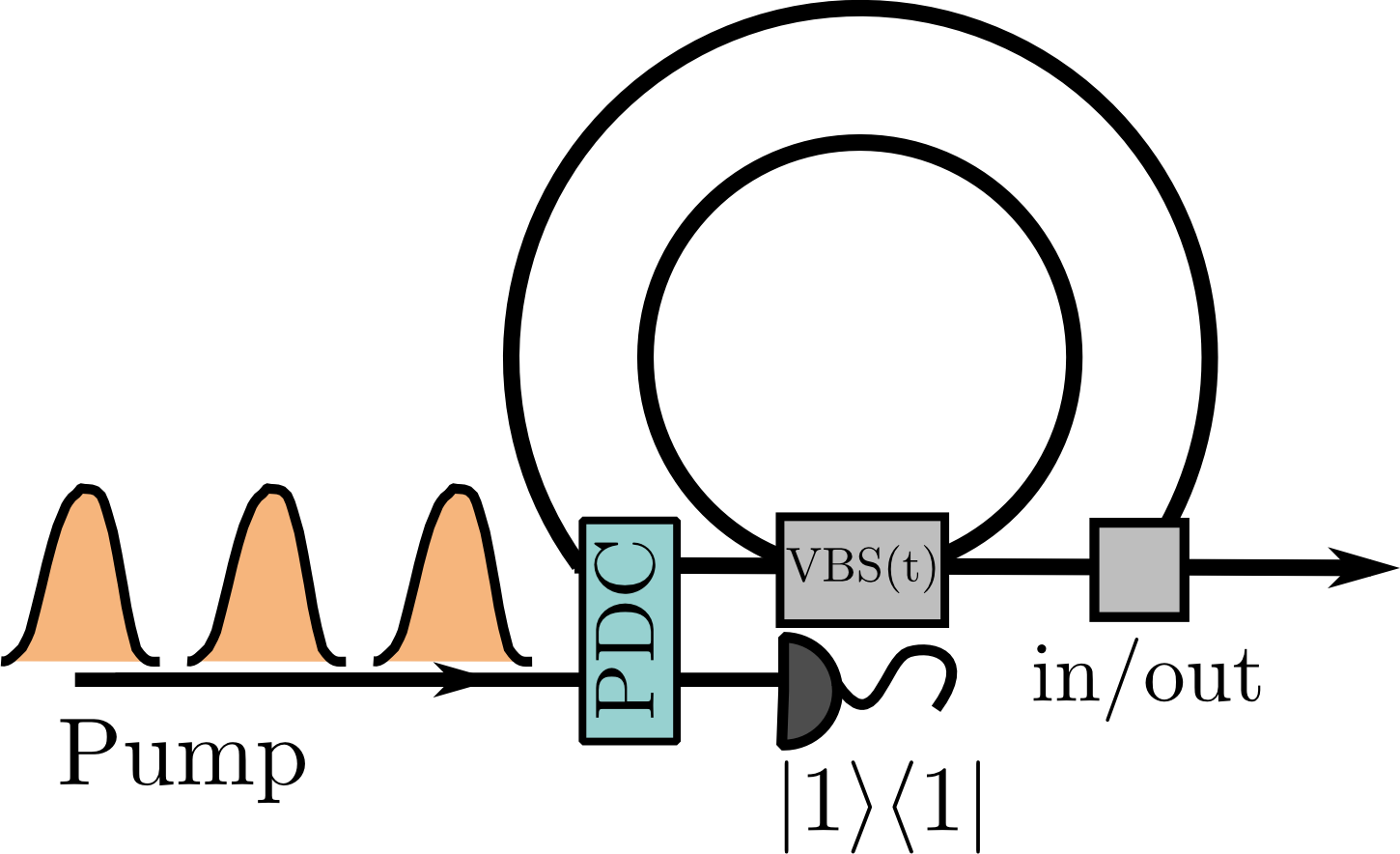}
 \caption{\label{fig:TM_Setup} Scheme of the time-multiplexing setup with a single photon (PDC) source within the loop structure, adapting the time-multiplexed scheme by~\cite{motes_scalable_2014}. The variable beam splitter VBS(t) implements all nodes in the network.}
\end{figure}

Perhaps more significantly than a constant factor speed-up is the dramatic reduction in the required optimal squeezing parameter $\lambda$ to achieve this improvement [Fig.~\ref{fig:probs} (b)]. 
By choosing $k=\sqrt{m}=n$, the optimal $\lambda$ reduces by $\lesssim 2$ orders of magnitude. 
Not only does this dramatically reduce pump power requirements for the $n^3$ sources, but also the probability of generating higher-order terms which act as noise sources on the signal, from which we can calculate the signal-to-noise ratio (SNR) (see supplementary material for further details). 
This means that large numbers of heralded single photons can be generated whilst higher-order contributions are almost completely suppressed, thus overcoming the need for photon-number-resolving detectors. 
Indeed, the SNR for DBS actually increases with photon number [Fig.~\ref{fig:probs} (c)], which makes PDC sources used in this manner a promising candidate for scaling up boson sampling experiments.
Although DBS significantly improves the measurement rate of a boson sampling experiment, this is at a cost of increased input resources $s$ (from originally $s=n$ sources to $s=m$ sources in SBS, to $s=k\cdot m$ sources in DBS). The operation of many sources of indistinguishable photons is a challenging task. However, employing techniques from time-multiplexed quantum networks~\cite{schreiber_photons_2010} inherits all the benefits of photon indistinguishability and homogeneity whilst reducing the physical overhead to a single set of components. Indeed, within the context of boson sampling, such a loop architecture has been proposed~\cite{motes_scalable_2014} and experimentally demonstrated~\cite{he_scalable_2016}. DBS is easily adapted to this approach by placing a down-conversion source within the loop structure, as shown in Fig.~\ref{fig:TM_Setup}, which is readily implemented with current technology.

In conclusion, we propose driven boson sampling to improve the generation rate of valid input states while reducing the necessary pump powers per source significantly. 
The reduction of pump power drastically decreases the impact of higher-order photon contributions and improves the SNR, demonstrating our approach as a promising candidate to scale up boson sampling machines. 

In line with previous approaches, we considered the generation (and detection) of \textit{exactly} $n$ single photons. 
However, as a final remark, it may be interesting to investigate whether the computational complexity of boson sampling remains if we sample the aggregate distribution of several boson sampling experiments, each with different (but well-defined) number.  
One specific example would be to consider the computational complexity of sampling the bosonic distribution when \textit{at least} $n$ photons are generated (and detected) from $q\geq n$ possible sources. Of course, were this computationally hard, this would further increase the input state generation efficiency with probabilistic, heralded single photons sources dramatically.

\textbf{Acknowledgements}
S.B. and T.J.B contributed equally to this work. This work has received funding from the European Union’s Horizon 2020 research and innovation programme under the QUCHIP project GA no. 641039 and by the DFG (Deutsche Forschungsgemeinschaft) via the Gottfried Wilhelm Leibniz-Preis.
T.J.B. acknowledges financial support from the DFG under SFB/TRR 142. 
C.S.H. and I.J. received financial support from Grants No. RVO 68407700 and No. GA\v{C}R 13-33906 S. 
The authors acknowledge T. Weich for helpful discussions.
\bibliography{thispaper}

\end{document}


\widetext
\clearpage
\begin{center}
\textbf{\large Supplemental Material: Driven Boson Sampling}
\end{center}
\setcounter{equation}{0}
\setcounter{figure}{0}
\setcounter{table}{0}
\setcounter{page}{1}
\makeatletter
\renewcommand{\theequation}{S\arabic{equation}}
\renewcommand{\thefigure}{S\arabic{figure}}
\renewcommand{\bibnumfmt}[1]{[S#1]}
\renewcommand{\citenumfont}[1]{S#1}

\section{Section 1: State generation probabilities}
As stated in the main text, the probability of state generation in a scattershot boson sampling experiment comprising $n$ photons in $m$ modes is given by
\begin{equation}\label{eqn:probs}
P_s^{\mathrm{SBS}}\left(n\right)=\left(\begin{array}{c}m\\n\end{array}\right)P_1^nP_0^{m-n},
\end{equation}
where $P_1$ is the probability of a single-photon being generated and $P_0$ is the probability of generating vacuum. In parametric down-conversion, these probabilities are given by
\begin{align}
P_1=&\left(1-\lambda^2\right)\lambda^2~,\\
P_0=&\left(1-\lambda^2\right)~,
\end{align}
where $\lambda$ is the squeezing parameter, such that Eq.~\ref{eqn:probs} maybe simplified to
\begin{equation}
P_s^{\mathrm{SBS}}\left(n\right)=\left(\begin{array}{c}m\\n\end{array}\right)\left(1-\lambda^2\right)^m\lambda^{2n}
\end{equation}
To maintain consistency with Ref.~\cite{lund}, we assume that the number of modes $m=n^2$.

In driven boson sampling, the number of possible modes in which a source can fire is now $k\cdot m$. Note that while $k$ can be freely chosen, we chose $k=\sqrt{m}=n$ in our analysis. Thus the probability of $n$ sources producing single photons and $k\cdot m-n$ sources producing vacuum is 
\begin{align}
P_s^{\mathrm{DBS}}\left(n\right)=&\left(\begin{array}{c}k\cdot m\\n\end{array}\right)P_1^nP_0^{k\cdot m-n}\\
=&\left(\begin{array}{c}n^3\\n\end{array}\right)P_1^nP_0^{n^3-n}~.
\end{align}

In both cases (SBS and DBS), an event is considered when $n$ heralded single-photon sources fire. Fig.~5 (a) in the main text shows the maximum probability of this occurring as a function of the number of desired photons $n$, with the corresponding optimal value of $\lambda$ given by
\begin{align}\label{eqn:lamopt}
\lambda_\textrm{opt}=&\sqrt{\frac{n}{m+n}}\\
\lambda_\textrm{opt}^\textrm{(SBS)}=&\sqrt{\frac{n}{n^2+n}}\\
\lambda_\textrm{opt}^\textrm{(DBS)}=&\sqrt{\frac{n}{k\cdot n^2+n}}
\end{align}
in accordance with the supplementary material in Ref.~\cite{lund}. For $k>1$ the optimal squeezing parameter for DBS is lower than for SBS  as shown in Fig.~6~(b).

\subsection{Noise contribution  from higher-order terms}
The state $\rho_H$ that is produced when one of the $n$ herald detectors fires is a mixture of single photons plus higher-order terms
\begin{equation}
\rho_{H}=\frac{\left(1-\lambda^2\right)}{\lambda^2}\sum_{i=1}^\infty\lambda^{2i}\ket{i}\!\!\bra{i}~.
\end{equation}
The probability that, given a herald click, the heralded state is a single photon state is
\begin{equation}
P_{1|H}=\left(1-\lambda^2\right)~,
\end{equation}
therefore the probability that all $n$ heralded states are single photons, given $n$ heralds, is $P_{1|H}^n=\left(1-\lambda^2\right)^n$. Noise events occur when states other than single photons are heralded. The probability of at least one noise event given $n$ heralds is $1-P_{1|H}^n$, therefore we can define the signal to noise ratio (SNR) as the ratio of these probabilities, {i.e.}
\begin{equation}
\textrm{SNR}=\frac{P_{1|H}^n}{1-P_{1|H}^n}=\frac{\left(1-\lambda^2\right)^n}{1-\left(1-\lambda^2\right)^n}
\end{equation}
Substituting Eq.~\ref{eqn:lamopt} for $\lambda$, we can write a general form of the SNR as a function of possible input modes $k\cdot m$
\begin{equation}\label{eqn:SNR}
\textrm{SNR}=\frac{\left(\frac{k\cdot m}{k\cdot m+n}\right)^n}{1-\left(\frac{k\cdot m}{k\cdot m+n}\right)^n}~.
\end{equation}
In an experimental scenario, the desired signal to noise ratio can be obtained by choosing an appropriately large $k$. We note that, for $k=1$ (as is the case in SBS), the SNR cannot exceed unity for $m=n^2$. This is shown in Fig.~5 (c) in the main text. By solving Eq.~\ref{eqn:SNR}$\geq$1 for $m$, we find that $m\geq\frac{n}{\sqrt[n]{2}-1}$. 

Conversely, for $m=n^2$, the minimum number of generation layers $k$ must exceed $\tfrac{1}{n\left(\sqrt[n]{2}-1\right)}$. We choose $k=n$ and $m=n^2$, which satisfies $k\geq \tfrac{1}{n\left(\sqrt[n]{2}-1\right)} ~\forall n$. Equality specifies the minimum depth of the generation network.


\section{Section 2: Photon addition by linear optics}
\begin{figure}[bh!]
 \centering
\includegraphics{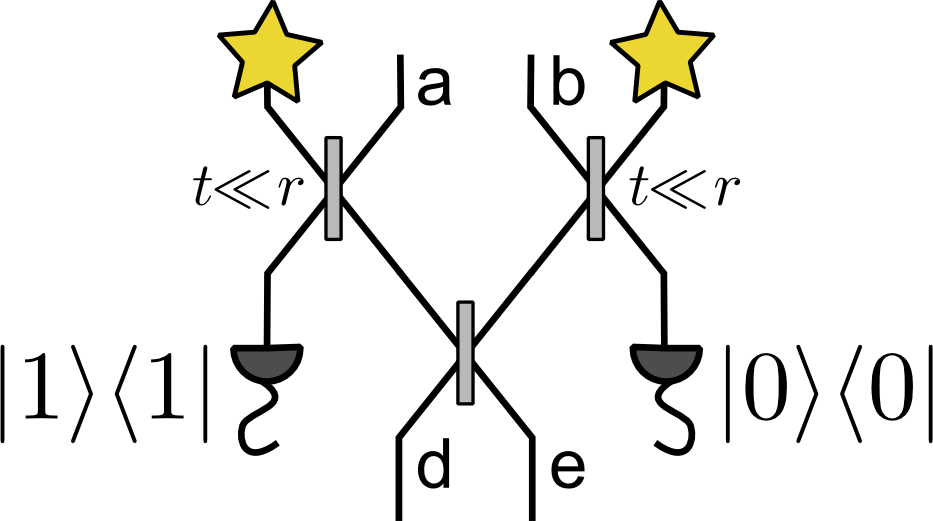}
 \caption{\label{fig:BS_icpl} Linear optical beam splitter equivalent of PDC incoupling process within the network}
\end{figure}

The experimental implementation of adding a photon to a particular mode is shown in Fig.~4 in the main text. 
In Fig.~\ref{fig:BS_icpl} we present an equivalent linear optics approach: In each network mode ($a$ and $b$), a source of single photons is incident on a highly-reflective beam splitter. Most of the time, the additional photon does not enter the network, while any existing photons in mode $\mathrm{b}$ are highly unlikely to couple out to the detection modes. In the rare cases that a detector registers vacuum, however, we can herald that a new photon has entered the network (in mode $b$ in this case). 
This process is difficult to implement with current technologies, since the generation and detection efficiencies must be very high to avoid errors.

